\documentclass{article}

\usepackage{amssymb}
\def\beq{\begin{equation}}
\def\eeq{\end{equation}}
\def\bce{\begin{center}}
\def\ece{\end{center}}
\def\bea{\begin{eqnarray}}
\def\eea{\end{eqnarray}}
\def\ben{\begin{enumerate}}
\def\een{\end{enumerate}}

\def\brr{\begin{array}}
\def\err{\end{array}}

\let\mbox=\hbox

\def\al{\alpha}
\def\be{\beta}
\def\ga{\gamma}
\def\de{\delta}

\def\ka{\kappa}
\def\la{\lambda}
\def\na{\nabla}

\def\si{\sigma}

\begin{document}

\begin{center}

{\Large {\bf On the cosmological effects of the Weyssenhoff spinning fluid
in the Einstein-Cartan framework}} \\
\vspace{3mm}

{\large Guilherme de Berredo-Peixoto} \\
\vspace{3mm}

Departamento de F\'{\i}sica, ICE, Universidade \\
Federal de Juiz de Fora, Campus universit\'ario \\
Juiz de Fora, MG 36036-330 Brazil \\
guilherme@fisica.ufjf.br
\vspace{3mm}

{\large Emanuel Antonio de Freitas} \\
\vspace{3mm}

Col\'egio T\'ecnico Universit\'ario, Universidade \\
Federal de Juiz de Fora, Campus universit\'ario, \\
Av. Bernardo Mascarenhas, 1283 - Juiz de Fora, MG Brazil 36080-001 \\
emanuel@fisica.ufjf.br

\end{center}

\begin{abstract}

The effects of non-Riemannian structures in Cosmology have been 
studied long ago and are still a relevant subject of investigation.
In the seventies, it was discovered that singularity avoidance 
and early accelerated expansion can be induced by torsion in
the Einstein-Cartan theory. In this framework, torsion is not
dynamical and is completely expressed by means of the spin 
sources. Thus, in order to study the effects of torsion in the
Einstein-Cartan theory, one has to introduce matter with spin.
In principle, this can be done in several ways. In this work we 
consider the cosmological evolution of the universe in the presence
of a constant isotropic and homogeneous axial current and the Weyssenhoff
spinning fluid. We analyse possible solutions of this model, with and 
without the spinning fluid. \\

keywords: Cosmology, Torsion, Einstein-Cartan theory, 
spinning fluid.

\end{abstract}


\section{Introduction}	

The non-Riemannian generalizations of General Relativity 
have been played an important role in gravity and particularly
in Cosmology. The most simple generalization of General
Relativity (preserving metricity) is achieved by the introduction of an asymetric connection, with torsion as its antisymmetric part. 
It is possible to consider
cosmological models in the framework of more general non-Riemannian
structures (see for example the review by Puetzfeld\cite{Puetzfeld}),
but these cases will not be treated here. For the introduction of the foundations of the theory, see Ref. \cite{revhehl}, and for
a more recent review, including the quantum aspects of torsion, 
see Ref. \cite{shapiro}. 

According to Kopczynski \cite{kopcz},
Trautman \cite{trautman} and Hehl, von der Heyde, Kerlick \cite{hehl}, 
the singularity avoidance and accelerated expansion can be induced 
by torsion in the Einstein-Cartan theory. In this framework, torsion is completely expressed in terms of the spin sources \cite{revhehl}, such that one has to introduce matter with spin in the gravitational 
action. Spin sources can be introduced, for example, by means of the
Dirac action, in the light of the variational principle (See, e.g., 
Ref. \cite{kerlick}, where torsion does not 
prevent the initial singularity, but rather enhances it). 

Alternatively, one can consider a fluid with intrinsic 
spin, not admitting {\it a priori} a Lagrangian full description. One
has to postulate a spin correction to the usual energy-momentum 
tensor. For example, Szydlowski and Krawiec \cite{krawiec} have studied the cosmological
effects of an exotic perfect fluid known as the Weyssenhoff
fluid \cite{weyssenhoff}, as well
as the constraints from supernovae Ia type observations, 
concluding that the dust Weyssenhoff fluid 
provides accelerated expansion but it 
can not serve as an
alternative to Dark Energy. Gasperini \cite{gasperini} considered the
Weyssenhoff fluid with its energy momentum tensor (derivable from a 
Lagrangian) improved by Ray and Smalley \cite{ray}, with spin as a thermodynamical variable. Obukhov and Korotky \cite{obukhov} formulated 
a more general variational theory describing the Weyssenhoff fluid and
also applied to cosmological models with rotation, shear and 
expansion.

In Ref. \cite{gasperini}, torsion provides 
singularity avoidance and inflation, but the expansion factor of the 
cosmological scale, $a(t)$, is too small, unless the state equation parameter $w$ ($p = w \rho$) of the spin fluid is fine tunned 
in a very special way. In this work we shall consider the 
Weyssenhoff fluid with the energy-momentum tensor improved by
Ray and Smalley \cite{ray}, in the relativistic regime, and also a constant axial current coupled
with torsion, $J^\mu S_\mu$ ($S_\mu =$ totaly antisymmetric torsion). 

\section{Variational Principle}

The action in the Einstein-Cartan framework is given by
\beq
S = \int \sqrt{-g} d^4 x\left\{ -\frac{1}{\ka^2} \tilde{R} + 
{\cal L}_M\right\}\, , \label{general action}
\eeq
where metric has signature ($+ - - -$), $\ka^2 = 16 \pi G$ (we use units
such that $\hbar = c = 1$) and $\tilde{R}$ is the Ricci scalar
constructed with the asymmetric connection,\footnote{All 
quantities with an upper tilde are 
constructed with the asymmetric connection,
and the corresponding quantities without tilde are constructed with
the Riemannian (symmetric) conection.} 
$\tilde{\Gamma}^\mu\mbox{}_{\al\be}$, which, by using the metricity
condition ($\tilde{\na}_\al g_{\mu\nu} = 0$) and the following definition 
of torsion 
$$
T^\mu\mbox{}_{\al\be} := \tilde{\Gamma}^\mu\mbox{}_{\al\be} - 
\tilde{\Gamma}^\mu\mbox{}_{\be\al}\, ,
$$
can be expressed as 
\beq
\tilde{\Gamma}^\mu\mbox{}_{\al\be} = 
\Gamma^\mu\mbox{}_{\al\be} + K^\mu\mbox{}_{\al\be}\, ,
\eeq
where $\Gamma^\mu\mbox{}_{\al\be}$ is the Riemannian connection
(Levi-Civita connection) and
the quantity $K^\mu\mbox{}_{\al\be}$ is the contortion tensor,
given by
$$
K^\mu\mbox{}_{\al\be} = \frac{1}{2}\left( T^\mu\mbox{}_{\al\be}
-T_\al\mbox{}^\mu\mbox{}_\be - T_\be\mbox{}^\mu\mbox{}_\al 
\right)\, .
$$
The term ${\cal L}_M$ is the Lagrangian describing matter distribution. 
We consider here the following matter Lagrangian:
\beq
{\cal L}_M = {\cal L}_{AC} + {\cal L}_{SF}\, ,
\eeq
where ${\cal L}_{SF}$ is the Lagrangian of spin fluid \cite{ray}
and ${\cal L}_{AC}$ is the external source, present in the minimally 
coupling Dirac sector (see, e.g., Ref. \cite{shapiro}):
\beq
{\cal L}_{AC} = J^\mu S_\mu\, ,
\eeq 
where $J^\mu$ is a constant background axial 
current\footnote{Here we do not consider $J^\mu
= \bar{\psi}\ga^5\ga^\mu\psi$ {\it a priori}.}
and $S_\mu$
is the axial part of torsion, defined by $S_\mu = \varepsilon_{\la\rho\si\mu}
T^{\la\rho\si}$ ($\varepsilon_{\la\rho\si\mu}$ is the Levi-Civita tensor, with $\sqrt{-g}\varepsilon_{0123} = 1$).

In order to vary the action and get the dynamical equations, let us
choose $g^{\mu\nu}$ and $T^\al\mbox{}_{\be\ga}$ as independent 
dynamical variables (as was done in Refs. 
\cite{revhehl,gasperini}), and $J^\mu$ 
as an external quantity. 

The algebraic equation for torsion, coming from variation with 
respect to $T^\al\mbox{}_{\be\ga}$, reads
\beq
T^{\mu\al\be} = -\ka^2\left\{  
2J^\si\varepsilon^{\mu\al\be}\mbox{}_\si + 
\frac12 S^{\al\be}u^\mu  \right\}\,, \label{T}
\eeq
where $S^{\al\be}$ is the spin tensor,
we have used\cite{gasperini,ray}
$$
\tau^{\mu\nu\al}_{SF} = \frac{1}{\sqrt{-g}}\frac{\de (\sqrt{-g}{\cal L}_{SF})}{\de K_{\mu\nu\al}} =
\frac12 S^{\mu\nu}u^\al\, ,
$$
and $u^\al$ is the fluid four-velocity. Substitution of Eq.
(\ref{T}) into the dynamical equations coming from variation with 
respect to $g^{\mu\nu}$ yields, after taking the average,
\bea
G_{\mu\nu} & = & 
\ka^4\left\{
-3g_{\mu\nu}J^2 - 6J_\mu J_\nu 
+ \frac{1}{16}g_{\mu\nu}\si^2 - 
\frac18 u_\mu u_\nu \si^2\right\} + g_{\mu\nu} \Lambda
\nonumber \\
& + & \frac{\ka^2}{2} \left\{
(\rho + p)u_\mu u_\nu - p g_{\mu\nu}\right\}\,,
\label{equations}
\eea
where $2\si^2 = <S_{\mu\nu}S^{\mu\nu}>$, $J^2 = J_\mu J^\mu$ and
we have used the same energy-momentum tensor for the spinning fluid
of Gasperini \cite{gasperini} and included the cosmological
constant, $\Lambda$.

\section{Dynamical Equations and Solutions}

The spacetime metric 
is the spatially flat homogeneous and isotropic metric,
$ds^2 = dt^2 - a(t)^2 (dx^2+dy^2+dz^2)$.
For the radiation fluid ($p = \rho/3$), the relevant components 
of Eqs. (\ref{equations}) can be used to get
\beq
\frac{\ddot{a}}{a} = \ka^4
\si^2/24 - \frac{\ka^2}{6}\rho +\Lambda/3\, .
\eeq
It is remarkable that the axial current has no effect in
the above equation.
Further manipulations give the energy conservation, 
$\dot{\rho}a + 4\dot{a} \rho = 0$, with solution coinciding
with the standard radiation dominated solution, 
$\rho = \rho_0\frac{a_0^4}{a^4}$ (subscript $0$ means present 
time). In order to find solution for $a(t)$, one substitutes
this result and the assumption\cite{ponomariev} 
$\si^2 = \ga\rho^{3/2}$ 
into the temporal component of Eqs. (\ref{equations}),
\beq
\frac{3\dot{a}^2}{a^2} = 
\ka^4\left(
-9J^2 - \si^2/16 \right) + \ka^2\rho/2 + \Lambda\, .
\eeq 

Let us choose $\rho_0 = 10^{-54}$ GeV$^4$ and keep in mind
$\Lambda = 5\times 10^{-84}$ GeV$^2$ and $\ka^2 = 
3.38\times 10^{-37}$ GeV$^{-2}$.
In the absence of the spin fluid, positivity of
$\dot{a}^2$ establishes the upper bound 
$J^2_0\simeq 4.863\times 10^{-12}$ GeV$^6$. 
For $J^2 < J^2_0 $ and $\ga = 0$, there is no singularity 
avoidance and late accelerated expansion begins at 
$a = 0.0136$ (independently on the $J^2$).
Similar upper bound occurs 
if we include the spin fluid.
For $\ga = 10^{-12}$, the minimum allowed
$a$ is $6.50\times 10^{-39}$, the (very brief) early 
accelerated expansion ends at $a = 9.19\times 10^{-39}$
and late accelerated expansion starts at $a = 0.0136$.
Actually, these results corresponds to the case $J^2 = 0$
($J^2 < J^2_0$ is still too small to give substantial effect). 
The two different accelerated expansion epochs is guaranteed
as far as $0 < \ga < 8\times 10^{59}$.

It is worth mentioning that Kosteleck\' y, Russell and Tasson
\cite{russell} have applied the recent experimental searches 
to find constraints for the constant torsion field. For example,
if torsion is minimally coupled with fermions, it is showed
(in our notations) that 
$|S_\mu S^\mu| < 8.4\times 10^{-54}$ GeV$^2$. Of course it does
not apply to the present consideration, because torsion depends also
on the spin density and so it is not constant. But in the absence of
the spin fluid, torsion is entirely written in terms of the axial 
current, such that it is constant. In this case, the above upper 
bound can be used to fix 
$J^2 \lessapprox 10^{18}$ GeV$^6$, which is remarkably much larger
than the upper bound dictated by positivity of $\dot{a}^2$. 

The investigation for
time-dependent $J^2$ is postponed for a forthcoming work.

\section*{Acknowledgments}
The authors are grateful to Prof. I.L. Shapiro for fruitful 
discussions. The authors also acknowledge support from FAPEMIG, and
G.B.P. is grateful to CNPq and FAPES for financial support.

\end{document}